\documentclass[aps,preprint,tightenlines]{revtex4}%
\usepackage{amsfonts}
\usepackage{amsmath}
\usepackage{amssymb}
\usepackage{graphicx}%
\begin{document}
\title[EIT atom squeezing]{Atom spin squeezing in a double $\Lambda$ system}
\author{A. Dantan, M.\ Pinard}
\affiliation{4 place Jussieu, F75252\ Paris Cedex 05 France}
\author{P. R. Berman}
\affiliation{Michigan Center for Theoretical Physics, FOCUS Center, and Physics Department,
University of Michigan, Ann Arbor, Michigan 48109-1120}

\begin{abstract}
The interaction of classical and quantized electromagnetic fields with an
ensemble of atoms in an optical cavity is considered. Four fields drive a
double-$\Lambda$ level scheme in the atoms, consisting of a pair of $\Lambda$
systems sharing the same set of lower levels. Two of the fields produce
maximum coherence, $\rho_{12}\approx-1/2$, between the ground state sublevels
$1$ and $2$. This \textit{pumping scheme }involves equal intensity fields that
are resonant with both the one and two-photon transitions of the $\Lambda$
system. There is no steady-state absorption of these fields, implying that the
fields induce a type electromagnetically-induced transparency (EIT) in the
medium. An additional\ pair of fields interacting with the second $\Lambda$
system, combined with the EIT\ fields, leads to squeezing of the atom spin
associated with the ground state sublevels. Our method involves a new
mechanism for creating steady-state spin squeezing using an optical cavity. As
the cooperativity parameter $C$ is increased, the optimal squeezing varies as
$C^{-1/3}$. For experimentally accessible values of $C$, squeezing as large as
90\% can be achieved.

\end{abstract}
\volumeyear{year}
\volumenumber{number}
\issuenumber{number}
\eid{identifier}
\date[Date text]{date}
\received[Received text]{date}

\revised[Revised text]{date}

\accepted[Accepted text]{date}

\published[Published text]{date}

\startpage{1}
\endpage{ }
\maketitle
\tableofcontents

\section{Introduction}

Spin squeezing refers to the reduction of noise in one of the components of
the effective spin associated with an ensemble of two-level quantum systems.
There has been a great deal of interest in spin squeezing as a means for
reducing the quantum noise that is intrinsic to any precision measurement
\cite{Wineland}. Several methods for achieving spin squeezing have been
proposed \cite{Polzik1,mandel,Vernac,Vernac2,Vernac3,bouchoule,andre,molmer},
including a recent one of ours involving atoms interacting with a classical
field and a quantized, cavity field in a $\Lambda$ configuration
\cite{dantan}. We have shown that, for a sufficiently large nonlinearity
(large cooperativity parameter $C$), \textit{self-squeezing} is obtained in
such a $3$-level atomic medium when the input field driving the cavity mode is
a coherent state of the radiation field. Maximal squeezing occurs near the
points of optical bistability in this system. With increasing cooperativity
parameter $C$, the maximum self-squeezing that can be obtained is about
$30\%$. In this paper, we show that it is possible to increase this limit by
modifying the pumping scheme. In effect, we introduce a method for pumping the
coherence between the two ground state levels using a double-$\Lambda$ scheme
\cite{double}.

The double $\Lambda$ scheme consists of a pair of $\Lambda$ systems sharing
the same set of lower levels. One of the $\Lambda$ schemes is designed to
produce maximum coherence, $\rho_{12}=-1/2$, between the ground state levels
$1$ and $2$. This \textit{pumping scheme }involves equal intensity fields that
are resonant with both the one and two-photon transitions of the $\Lambda$
system and leads to pumping of the dark-state, $\left[  \left(  \left\vert
1\right\rangle -\left\vert 2\right\rangle \right)  /\sqrt{2}\right]  $. There
is no steady-state absorption of these fields, implying that the fields induce
a type electromagnetically-induced transparency (EIT)
\cite{hau,phillips,scully}. The EIT fields' quantum correlations have been
investigated in \cite{nussenzveig2}, but not the atom-field correlations. It
is not difficult, however, to show that, if the input fields are in a coherent
state, the spin associated with levels $1$ and $2$ resulting from these fields
alone is at least as noisy that of a coherent spin state. Hence, EIT alone
does not produce atomic spin squeezing. In order to squeeze the ground state
atomic spin, we introduce an asymmetry in the system by considering an
additional\ pair of fields interacting with the second $\Lambda$ system
\cite{dantan}. In the combined double-$\Lambda$ scheme, our calculations
predict that there is no limit to the spin squeezing that can be achieved; the
variance of one spin component approaches zero asymptotically as $C^{-1/3}$
for $C\gg1$. We provide analytical calculations of the atomic variance,
allowing one to optimize the squeezing and also discuss the process leading to
the creation of squeezing, which is new and different from squeezing
originating from optical bistability \cite{Vernac,Vernac3}. The EIT
interaction enables one to pump the coherence and increase the spin mean value
while the cavity coupling keeps the fluctuations low.

In Sec. II, we describe the system and give the set of Heisenberg-Langevin
equations governing the system. In Sec. III, we provide an effective $2$-level
system, give analytical results for the optimal squeezing and discuss the
squeezing creation process.

\section{Atom-field configuration}

\begin{figure}
[ptb]
\begin{center}
\includegraphics[
natheight=3.377100in,
natwidth=3.717000in,
height=2.9611in,
width=3.2578in
]%
{../../Paper 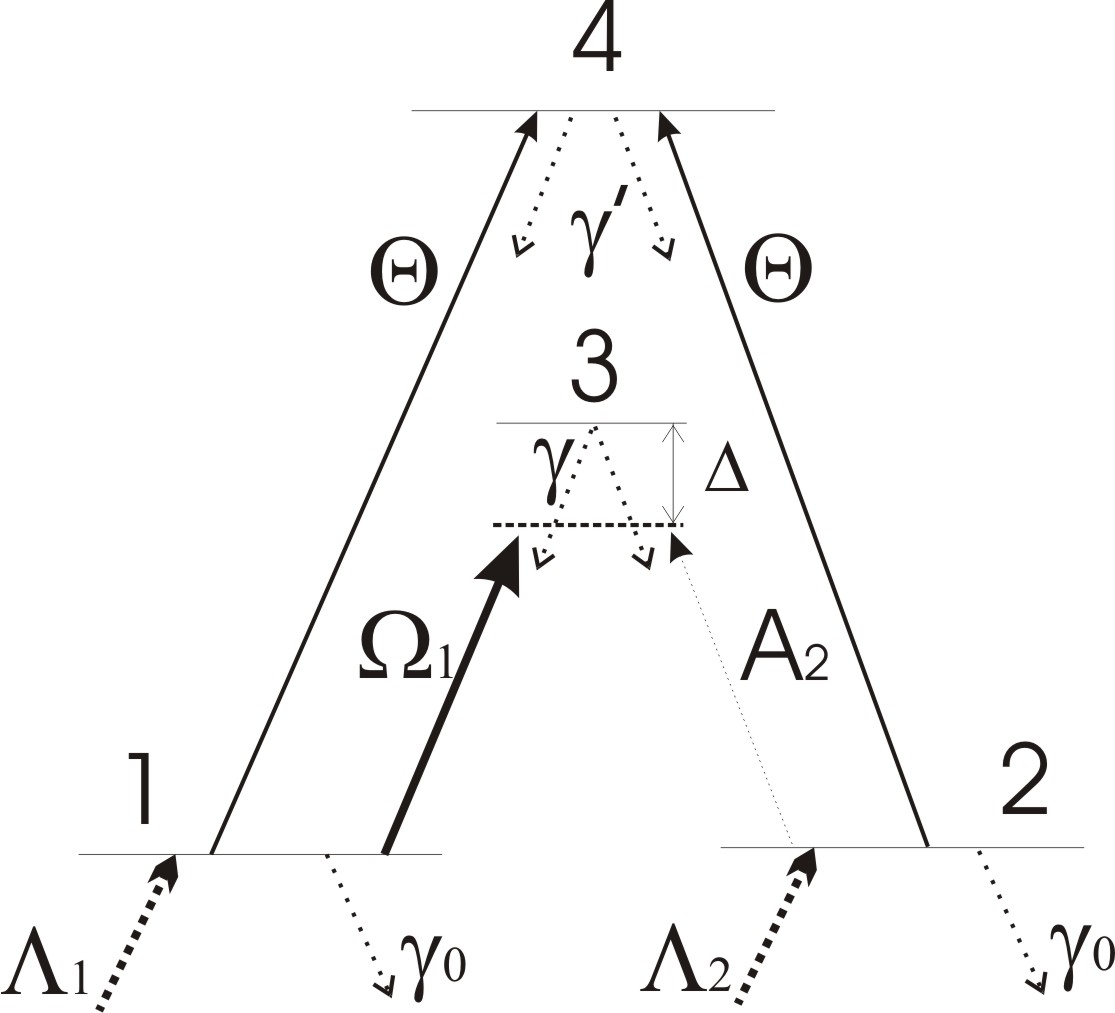}%
\caption{Double-$\Lambda$ scheme.}%
\label{Fig. 1}%
\end{center}
\end{figure}

The system considered in this paper consists of a set of $N$, $4$-level atoms,
whose levels form a double-$\Lambda$ configuration, as represented in Fig.
\ref{Fig. 1}.%
On the lower $\Lambda$ transition (1,3,2) the atoms interact with two light
fields: an intense classical field $A_{1}$ in a single pass scheme on
transition $1\rightarrow3$, and a quantum field $A_{2}$ in an optical cavity
on the transition $2\rightarrow3$. The field frequencies are $\omega_{1}$ and
$\omega_{2}$ and the detunings from atomic resonance $\Delta_{i}=\omega
_{3i}-\omega_{i}$ ($i=1,2$) are assumed to be much greater than the excited
decay rate of state $3$. If $\omega_{c}$ is the cavity resonance frequency
that is closest to the probe frequency, the cavity detuning for the quantum
field can be defined as $\Delta_{c}=$ $\omega_{2}-\omega_{c}.$ An incoming
quantum field $A_{2}^{in}$ drives the cavity field $A_{2}$. The field $A_{1}$
is treated classically and its intensity is supposed to be much greater than
that of the quantum field. On the second $\Lambda$ transition, the atoms
resonantly interact with the two modes of a quantum cavity field $\Theta$,
with frequency $\omega^{\prime}$ and cavity detuning $\Delta_{c}^{\prime}$.
Although both modes are represented by the same symbol $\Theta$, each mode
drives only one transition, $1\rightarrow4$ or $2\rightarrow4$. This
selectivity can be provided by field polarization if states $1$ and $2$ are
degenerate, or by frequency selectivity if they belong to different ground
state hyperfine manifolds.

The $4$-level system is described using $16$ collective operators for the $N$
atoms of the ensemble: the populations $\Pi_{i}=\sum\limits_{\mu=1}%
^{N}\left\vert i\right\rangle _{\mu}\left\langle i\right\vert _{\mu}$
($i=1-4$), the components of the optical dipoles $P_{ij}$ in the frames
rotating at the frequency of their corresponding lasers and their hermitian
conjugates and the components of the dipole associated to the ground state
coherence: $P_{r}=\sum\limits_{\mu=1}^{N}\left\vert 2\right\rangle _{\mu
}\left\langle 1\right\vert _{\mu}$ and $P_{r}^{\dagger}$. We take $\omega
_{1}=\omega_{2}$, so that the ground state coherence is excited at zero
frequency in both $\Lambda$s.

The coupling constant between atoms and field $A_{2}$ is defined by
$g=\mathcal{E}_{0}d/\hbar$, where $d$ is the atomic dipole, and $\mathcal{E}%
_{0}=\sqrt{\hbar\omega_{2}/2\epsilon_{0}\mathcal{S}c}$. With this definition,
the mean square value of the field is expressed in number of photons per
second. A second coupling constant $g^{\prime}$ is similarly defined for field
$\Theta$. The decay constants of dipoles $P_{13}$ and $P_{23}$ are taken equal
to $\gamma,$ and those of $P_{14}$ and $P_{24}$ equal to $\gamma^{\prime}$. In
order to take into account the finite lifetime of the two fundamental
sublevels $1$ and $2$, we include in the model another decay rate $\gamma_{0}%
$, which is supposed to be much smaller than $\gamma$. For example $\gamma
_{0}^{-1}$ can represent an atom's transit time in the light field, typically
of the order of a few milliseconds for cold atoms. On the other hand, $\gamma$
and $\gamma^{\prime}$\ are of the order of the MHz for excited states. We also
consider that the sublevels $1$ and $2$ are repopulated with incoherent
pumping terms $\Lambda_{1}$ and $\Lambda_{2}$, so that the total atomic
population is kept constantly equal to $N$.

The system evolution is given by a set of quantum Heisenberg-Langevin equations%

\begin{align}
\frac{d\Pi_{1}}{dt}  &  =i\Omega_{1}^{\ast}P_{13}-i\Omega_{1}P_{13}^{\dagger
}+ig^{\prime}\Theta^{\dagger}P_{14}-ig^{\prime}\Theta P_{14}^{\dagger}%
+\gamma\Pi_{3}+\gamma^{\prime}\Pi_{4}-\gamma_{0}\Pi_{1}+\Lambda_{1}%
+F_{11}\label{pi1}\\
\frac{d\Pi_{2}}{dt}  &  =igA_{2}^{\dagger}P_{23}-igA_{2}P_{23}^{\dagger
}+ig^{\prime}\Theta^{\dagger}P_{24}-ig^{\prime}\Theta P_{24}^{\dagger}%
+\gamma\Pi_{3}+\gamma^{\prime}\Pi_{4}-\gamma_{0}\Pi_{2}+\Lambda_{2}%
+F_{22}\label{pi2}\\
\frac{d\Pi_{3}}{dt}  &  =-(i\Omega_{1}^{\ast}P_{13}-i\Omega_{1}P_{13}%
^{\dagger})-(igA_{2}^{\dagger}P_{23}-igA_{2}P_{23}^{\dagger})-2\gamma\Pi
_{3}+F_{33}\label{pi3}\\
\frac{d\Pi_{4}}{dt}  &  =-(ig^{\prime}\Theta^{\dagger}P_{14}-ig^{\prime}\Theta
P_{14}^{\dagger})-(ig^{\prime}\Theta^{\dagger}P_{24}-ig^{\prime}\Theta
P_{24}^{\dagger})-2\gamma^{\prime}\Pi_{4}+F_{44}\label{pi4}\\
\frac{dP_{13}}{dt}  &  =-(\gamma+i\Delta_{1})P_{13}+i\Omega_{1}(\Pi_{1}%
-\Pi_{3})+igA_{2}P_{r}^{\dagger}+F_{13}\label{p13}\\
\frac{dP_{23}}{dt}  &  =-(\gamma+i\Delta_{2})P_{23}+igA_{2}(\Pi_{2}-\Pi
_{3})+i\Omega P_{r}+F_{23}\label{p23}\\
\frac{dP_{14}}{dt}  &  =-\gamma^{\prime}P_{14}+ig^{\prime}\Theta(\Pi_{1}%
-\Pi_{4})+ig^{\prime}\Theta P_{r}^{\dagger}+F_{14}\label{p14}\\
\frac{dP_{24}}{dt}  &  =-\gamma^{\prime}P_{24}+ig^{\prime}\Theta(\Pi_{2}%
-\Pi_{4})+ig^{\prime}\Theta P_{r}+F_{24}\label{p24}\\
\frac{dP_{r}}{dt}  &  =-\left(  \gamma_{0}-i\delta\right)  P_{r}+i\Omega
_{1}^{\ast}P_{23}-igA_{2}P_{13}^{\dagger}+ig^{\prime}\Theta^{\dagger}%
P_{24}-ig^{\prime}\Theta P_{14}^{\dagger}+F_{21}\label{pr}\\
\frac{dA_{2}}{dt}  &  =-(\kappa+i\Delta_{c})\text{ }A_{2}+\frac{ig}{\tau
}P_{23}+\sqrt{\frac{2\kappa}{\tau}}A_{2}^{in}\label{A2}\\
\frac{d\Theta}{dt}  &  =-(\kappa^{\prime}+i\Delta_{c}^{\prime})\text{ }%
\Theta+\frac{ig^{\prime}}{\tau^{\prime}}(P_{14}+P_{24})+\sqrt{\frac
{2\kappa^{\prime}}{\tau^{\prime}}}\Theta^{in} \label{theta}%
\end{align}
where $g$ and $g^{\prime}$ are assumed real, $\Omega_{1}=gA_{1}$,
$\delta=\Delta_{1}-\Delta_{2}$ is the detuning between the ground state
sublevels, $\kappa$ and $\kappa^{\prime}$ are the intracavity field decays and
$\tau$ and $\tau^{\prime}$ are the round trip times in the cavity. The
EIT\ fields are taken to be quantum cavity fields in the bad cavity limit
($\gamma^{\prime}\ll\kappa^{\prime}$), but could just as well have been taken
to be classical fields propagating in free space. From the previous set of
equations, it is possible to derive the steady state values and the
correlation matrix for the fluctuations of the atom-field system (see e.g.
\cite{Vernac}). Our aim here is to obtain the fluctuations of the spin
operators associated with levels $1$ and $2$ from simplified equations for the
ground state variables as in Ref. \cite{dantan}.

\section{Simplified equations for the ground state observables}

Owing to the off-resonant interaction on transitions $1\rightarrow3$ and
$2\rightarrow3$, the excited state population $\left\langle \Pi_{3}%
\right\rangle $ is negligible and the optical coherences $P_{13}$ and $P_{23}$
evolve rapidly compared to $\Pi_{1},\Pi_{2}$ and $P_{r}$. It is also
reasonable to assume that $\delta=\Delta_{1}-\Delta_{2}\ll\Delta=(\Delta
_{1}+\Delta_{2})/2$. On transitions $1\rightarrow4$ and $2\rightarrow4$ we
choose a pumping rate $\Gamma_{p}^{\prime}=2g^{\prime2}\left\vert
\Theta\right\vert ^{2}/\gamma^{\prime}$ much smaller than $\gamma^{\prime}$ so
that $\left\langle \Pi_{4}\right\rangle $ is negligible and the optical
coherences $P_{14}$ and $P_{24}$ adiabatically follow the ground state
observables. We assume also that the EIT cavity is sufficiently bad to ensure
that there is no bistability of these fields. Eliminating the excited state
populations and replacing the optical coherences in Eqs. (\ref{pi1}-\ref{pr})
by their steady state values, one gets simplified equations for the ground
state variables $S_{+}=P_{r}$, $S_{-}=P_{r}^{\dagger}$, $S_{z}=(\Pi_{2}%
-\Pi_{1})/2$ and the quantum field $A_{2},$%

\begin{align}
\frac{dS_{+}}{dt}  &  =-(\tilde{\gamma}_{0}-i\tilde{\delta})S_{+}%
+\tilde{\Lambda}_{12}+2i\tilde{g}A_{2}S_{z}+F_{+}\label{S+}\\
\frac{dS_{z}}{dt}  &  =-\tilde{\gamma}_{0}S_{z}+\frac{\tilde{\Lambda}%
_{2}-\tilde{\Lambda}_{1}}{2}+i\tilde{g}(A_{2}^{\dagger}S_{+}-S_{-}A_{2}%
)+F_{z}\label{Sz}\\
\frac{dA_{2}}{dt}  &  =-(\kappa+i\Delta_{c})A_{2}+i\frac{\tilde{g}}{\tau}%
S_{+}+\sqrt{\frac{2\kappa}{\tau}}A_{2}^{in} \label{evA2}%
\end{align}
where $\tilde{\delta}=\delta+\frac{\left\vert \Omega_{1}^{2}\right\vert
}{\Delta}$ is the effective atomic detuning corrected with the light-shift and
$\tilde{g}=g\frac{\Omega_{1}}{\Delta}$ is the effective coupling constant.
Denoting by $\Gamma_{p}=\gamma\left\vert \Omega_{1}^{2}\right\vert /\Delta
^{2}$ the optical pumping rate due to field $A_{1}$, the new in-terms and
decay constants are then%

\begin{equation}
\tilde{\gamma}_{0}=\gamma_{0}+\Gamma_{p}+\Gamma_{p}^{\prime};\quad
\tilde{\Lambda}_{2}-\tilde{\Lambda}_{1}=\Lambda_{2}-\Lambda_{1}+N\Gamma
_{p};\quad\tilde{\Lambda}_{12}=-N\Gamma_{p}^{\prime}/2 \label{interms}%
\end{equation}

We assume a symmetrical configuration ($\Lambda_{1}=\Lambda_{2}$) and that the
population is constant ($\Lambda_{1}+\Lambda_{2}=N\gamma_{0}$). To get
(\ref{evA2}), we have used the fact that $\left\vert \Omega_{1}^{2}\right\vert
\gg\left\vert g^{2}A_{2}^{\dagger}A_{2}\right\vert $. This effective system is
now quite similar to that of Ref. \cite{dantan} in the case of the single
$\Lambda$ Raman interaction, the essential difference is that EIT or
dark-state pumping results in an in-term for the ground state coherence. As
usual, we introduce the parameter $C$ quantifying the cooperative behavior of
the atomic ensemble%

\begin{equation}
C=\frac{g^{2}N}{2\kappa\tau\gamma} \label{CtildeC}%
\end{equation}

The steady-state can be obtained setting the time derivatives to $0$ in Eqs.
(\ref{S+})-(\ref{evA2}) and using the fact that the Langevin operators mean
values are $0$. Since we are interested in the quantum fluctuations, we
linearize the effective equations around their steady-state values, assuming
fluctuations are small with respect to mean values.

\subsection{Linearization and diffusion matrix\label{diffusion}}

The linearized equations for the fluctuations may be written in a matrix form%

\begin{equation}
\frac{d\ \left\vert \delta\xi(t)\right]  }{dt}=-\left[  B\right]  \left\vert
\delta\xi(t)\right]  +\left\vert F_{\xi}\right]  \label{evolution}%
\end{equation}
where $\left\vert \delta\xi(t)\right]  $ is the fluctuation vector $\left\vert
\delta\xi(t)\right]  =\left[  \delta A_{2}(t),\delta A_{2}^{\dagger}(t),\delta
S_{+}(t),\delta S_{-}(t),\delta S_{z}(t)\right\vert ^{T}$, $\left[  B\right]
$ is the linearized evolution matrix%

\begin{equation}
\left[  B\right]  =\left(
\begin{array}
[c]{ccccc}%
\kappa+i\Delta_{c} & 0 & -i\tilde{g}/\tau & 0 & 0\\
0 & \kappa-i\Delta_{c} & 0 & i\tilde{g}/\tau & 0\\
-2i\tilde{g}\langle S_{z}\rangle & 0 & \tilde{\gamma}_{0}-i\tilde{\delta} &
0 & -2i\tilde{g}\langle A_{2}\rangle\\
0 & 2i\tilde{g}\langle S_{z}\rangle & 0 & \tilde{\gamma}_{0}+i\tilde{\delta} &
2i\tilde{g}\langle A_{2}\rangle^{\ast}\\
i\tilde{g}\langle S_{-}\rangle & -i\tilde{g}\langle S_{+}\rangle & -i\tilde
{g}\langle A_{2}\rangle^{\ast} & i\tilde{g}\langle A_{2}\rangle &
\tilde{\gamma}_{0}%
\end{array}
\right)  \label{matrixB}%
\end{equation}
and $\left\vert F_{\xi}\right]  $ is the column vector regrouping the
corresponding Langevin operators. As in \cite{Vernac}, we define the
covariance matrix $\left[  G(t)\right]  $ by%

\begin{equation}
\left[  G(t)\right]  =\left\vert \delta\xi(t)\right]  \left[  \delta
\xi(0)\right\vert \label{Gt}%
\end{equation}
and the diffusion matrix by%

\begin{equation}
\left|  F_{\xi}(t)\right]  \left[  F_{\xi}(t^{\prime})\right|  =\left[
D\right]  \text{ }\delta(t-t^{\prime}) \label{D}%
\end{equation}

The values of the atomic diffusion coefficients can be derived from the
quantum regression theorem \cite{Cohen}. The complete diffusion matrix is
given in Appendix. The variances of the spin components and their correlation
functions are the elements of the zero time correlation matrix $\left[
G(0)\right]  $, which satisfies \cite{Carmichael}%

\begin{equation}
\left[  B\right]  \left[  G(0)\right]  +\left[  G(0)\right]  \left[  B\right]
^{\dagger}=\left[  D\right]  \label{G0}%
\end{equation}
The inverse of Eq. (\ref{G0}) gives $\left[  G(0)\right]  ,$ and,
consequently, the spin variances. We then proceed with the calculation of the
minimal variance in the plane orthogonal to the mean spin as in
\cite{Vernac,dantan}.

\subsection{Optimal squeezing for $\left\langle A_{2}\right\rangle
=0$\label{variance}}

The optimal atomic squeezing was found to occur at two-photon resonance
($\tilde{\delta}=0$) and when the quantum cavity field mean value was $0$,
with no cavity detuning ($\tilde{\Delta}_{c}=0$). The effect of a non-zero
mean value of the cavity field will be discussed later. In particular, we will
show that the results obtained in this Section still hold for small values of
the quantum field intensity. Since the calculations are much easier and the
physical meaning quite clear, we focus on the case $\left\langle
A_{2}\right\rangle =0$. The steady state is then simple: $\left\langle
S_{z}\right\rangle =N\Gamma_{p}/2\tilde{\gamma}_{0}$ and $\left\langle
S_{+}\right\rangle =\left\langle S_{-}\right\rangle =-N\Gamma_{p}^{\prime
}/2\tilde{\gamma}_{0}$. Rewriting the equations for atomic fluctuations in the
$S_{x}$, $S_{y}$, $S_{z}$ basis and defining the usual quadrature operators
for $A_{2}$,%

\begin{equation}
E_{P}=\frac{A_{2}+A_{2}^{\dagger}}{2},\quad E_{Q}=\frac{A_{2}-A_{2}^{\dagger}%
}{2i} \label{quadratures}%
\end{equation}
one obtains the following set of equations%

\begin{align}
\frac{d}{dt}\delta S_{x}  &  =-\tilde{\gamma}_{0}\delta S_{x}-2\tilde
{g}\left\langle S_{z}\right\rangle \delta E_{Q}+F_{x}\label{deltaSx}\\
\frac{d}{dt}\delta S_{y}  &  =-\tilde{\gamma}_{0}\delta S_{y}+2\tilde
{g}\left\langle S_{z}\right\rangle \delta E_{P}+F_{y}\label{deltaSy}\\
\frac{d}{dt}\delta S_{z}  &  =-\tilde{\gamma}_{0}\delta S_{z}-2\tilde
{g}\left\langle S_{+}\right\rangle \delta E_{Q}+F_{z}\label{deltaSz}\\
\frac{d}{dt}\delta E_{P}  &  =-\kappa\delta E_{P}-\frac{\tilde{g}}{\tau}\delta
S_{y}+\sqrt{\frac{2\kappa}{\tau}}\delta E_{P}^{in}\label{deltaEp}\\
\frac{d}{dt}\delta E_{Q}  &  =-\kappa\delta E_{Q}+\frac{\tilde{g}}{\tau}\delta
S_{x}+\sqrt{\frac{2\kappa}{\tau}}\delta E_{Q}^{in} \label{deltaEq}%
\end{align}

\begin{figure}
[ptb]
\begin{center}
\includegraphics[
natheight=3.579500in,
natwidth=5.800300in,
height=3.627in,
width=5.8591in
]%
{../../Paper 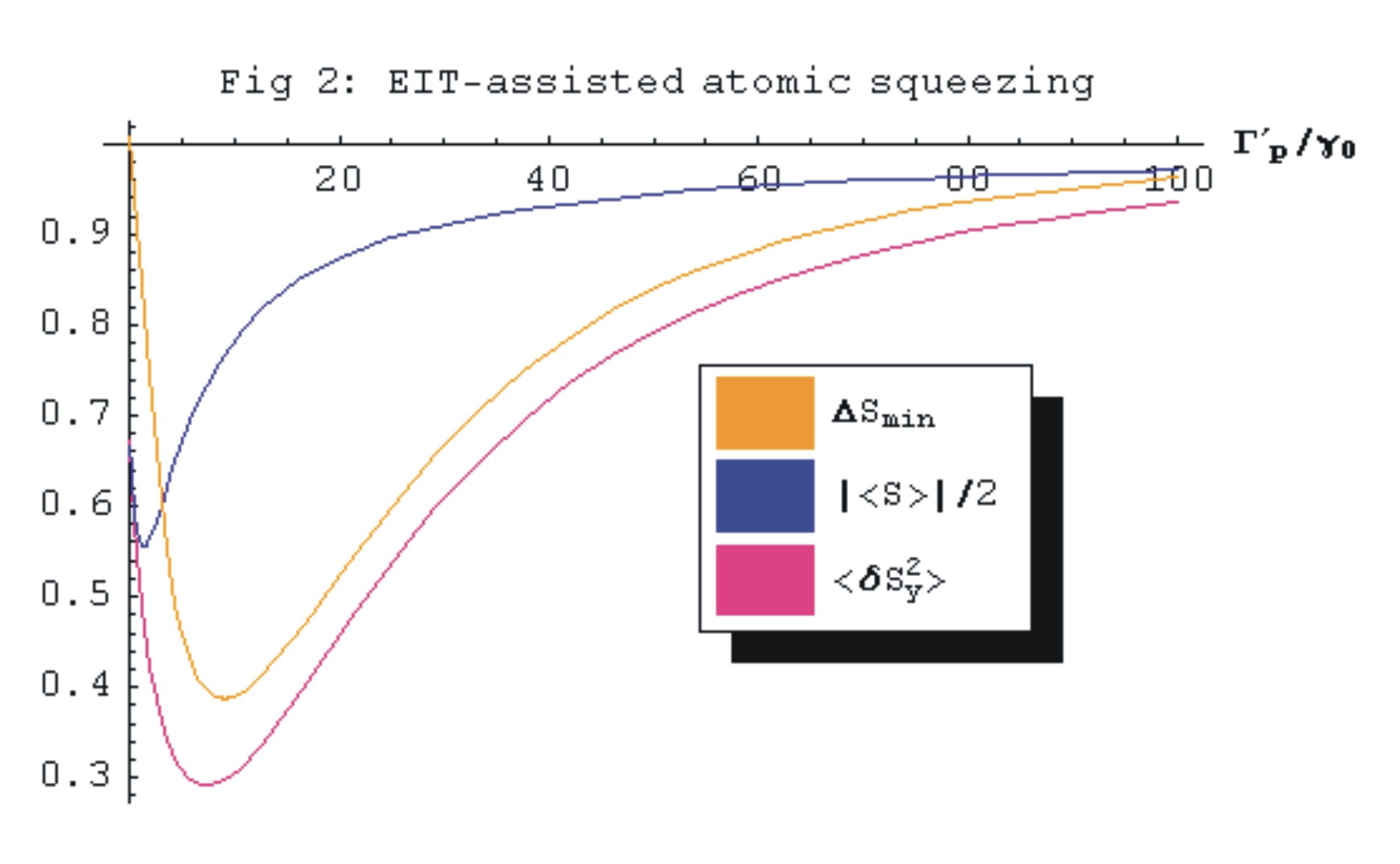}%
\caption{ Minimal variance $\Delta S_{\min}$ versus EIT pumping rate
$\Gamma_{p}^{\prime}$ (expressed in $\gamma_{0}$ units). The mean spin half
value $\left\vert \left\langle \mathbf{S}\right\rangle \right\vert /2$
and\ the minimal spin fluctuations $\left\langle \delta S_{y}^{2}\right\rangle
$\ (normalized by $N/4$) are also represented for the same parameters
($C=100$, $\rho=1/2000$, $\Gamma_{p}=2\gamma_{0}$).}%
\label{Fig. 2}%
\end{center}
\end{figure}

It can be seen that $\delta S_{y}$ is coupled only to $\delta E_{P},$ while
$\delta S_{x}$ and $\delta S_{z}$ are coupled together via $\delta E_{Q}.$ By
Fourier-transforming and integrating the linear set of equations, one can show
that the minimal variance is that of the $y$-component of the spin, which is
unchanged when one transforms to a basis where the z-axis is aligned along the
mean spin. The noisier component is then a linear combination of $\delta
S_{z}$ and $\delta S_{x}$. Explicitly, one obtains for the fluctuations of
$S_{y}$%

\begin{equation}
\left\langle \delta S_{y}^{2}\right\rangle =\frac{N}{4}\left[  1-\frac
{2C}{1+\tilde{\rho}}\frac{\Gamma_{p}^{2}(\gamma_{0}+\Gamma_{p}^{\prime}%
)}{\tilde{\gamma}_{0}(\tilde{\gamma}_{0}^{2}+2C\Gamma_{p}^{2})}\right]
\label{fluctuationsSy}%
\end{equation}
in which $\tilde{\rho}=\tilde{\gamma}_{0}/\kappa=\rho$ $\tilde{\gamma}%
_{0}/\gamma_{0}$ is the ratio of atomic and field decay rate and $\rho
=\gamma_{0}/\kappa$. Equation (\ref{fluctuationsSy}) shows clearly that the
fluctuations are small\ when $\tilde{\rho}\ll1$ (bad-cavity limit) and $C\gg1$
(high cooperative behavior). The criterion for spin squeezing is obtained
comparing the minimal variance to half the spin mean value \cite{Wineland},
\begin{equation}
\left\vert \left\langle \mathbf{S}\right\rangle \right\vert /2=N\frac
{\sqrt{\Gamma_{p}^{\prime2}+\Gamma_{p}^{2}}}{4\tilde{\gamma}_{0}}.\label{spin}%
\end{equation}
The atoms are said to be squeezed when $\Delta S_{\min}=\ \left\langle \delta
S_{y}^{2}\right\rangle /(\left\vert \left\langle \mathbf{S}\right\rangle
\right\vert /2)<1$. In Fig. \ref{Fig. 2}, we plot the minimum variance as a
function of the EIT pumping rate: $\Delta S_{\min}$ goes through a minimum in
the range of pump strengths satisfying $\gamma_{0}\ll\Gamma_{p}^{\prime}%
\ll\gamma^{\prime}$. The fluctuations of (\ref{fluctuationsSy}) should be
compared to the mean spin half value $\left\vert \left\langle \mathbf{S}%
\right\rangle \right\vert /2$ of (\ref{spin}). These two quantities, normalize%
d by $N/4$, are plotted versus the EIT\ pumping rate in Fig. 2 for given
values of $C$, $\tilde{\rho}$ and $\Gamma_{p}$. Both go through a minimum with
increasing $\Gamma_{p}^{\prime}$, but there exists a regime in which the spin
mean value is increased more than the fluctuations. In this regime, the atom
spin is squeezed. Note that when the EIT interaction is absent ($\Gamma
_{p}^{\prime}=0$), as well as when it is predominant ($\Gamma_{p}^{\prime}%
\sim100\gamma_{0}$), there is little or no squeezing. However, for
intermediate values of the pumping rate, the EIT interaction allows one to
pump the mean spin while the fluctuations are close to their minimal value.
The best squeezing for a fixed value of $\Gamma_{p}$ is determined by the
biggest \textquotedblright gap\textquotedblright\ between $\left\langle \delta
S_{y}^{2}\right\rangle $ and $\left\vert \left\langle \mathbf{S}\right\rangle
\right\vert /2$. The qualitative dependence of spin squeezing can be
understood as follows: one must have $\Gamma_{p}^{\prime}\gg\gamma_{0}$ to
produce significant pumping of the coherence $\rho_{12}$; however, the
entanglement of the spins with the cooperativity parameter must be
sufficiently large to dominate the fluctuations produced by the pumping fields
- this translates into the conditions, $\Gamma_{p}^{\prime}\ll\kappa$,
$C\Gamma_{p}^{2}/\Gamma_{p}^{\prime2}\geq1$, which are violated for
sufficiently large $\Gamma_{p}^{\prime}$. Note also that for $\Gamma_{p}=0$
(EIT interaction alone), the fluctuations are $N/4$, and $\Delta S_{\min
}=1+\gamma_{0}/$ $\Gamma_{p}^{\prime}\geq1$ confirming our statement that EIT
alone does not produce atomic squeezing.

\subsection{Optimized variance}

\begin{figure}
[ptb]
\begin{center}
\includegraphics[
natheight=3.039800in,
natwidth=6.643500in,
height=3.0839in,
width=6.7075in
]%
{../../Paper 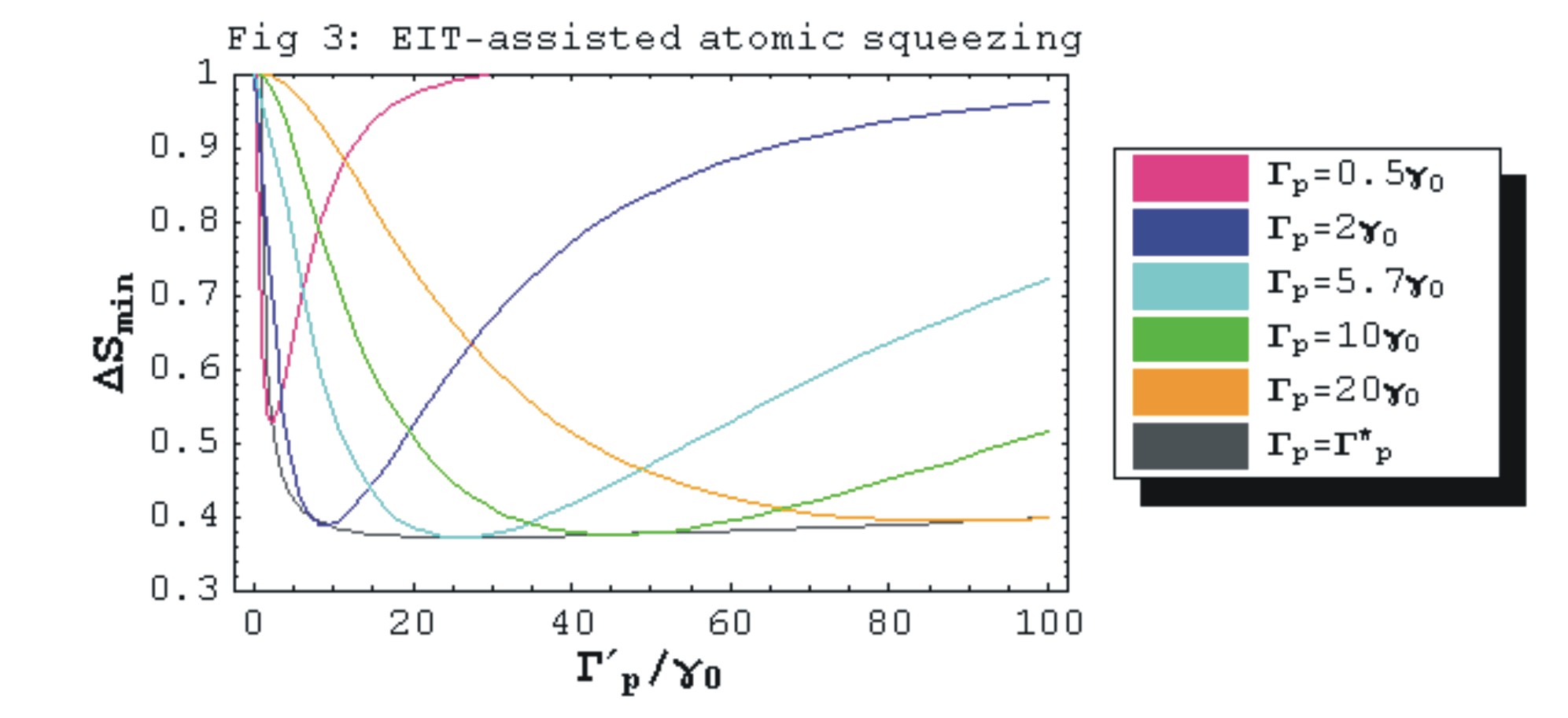}%
\caption{Minimal variance $\Delta S_{\min}$ versus $\Gamma_{p}^{\prime}$ (in
$\gamma_{0}$ units), for various values of the Raman pumping rate $\Gamma
_{p}=0.5,2,5.7,10,20$. $C$ and $\rho$ are equal to $100$ and $1/2000$. The
last curve shows the minimal variance in which $\Gamma_{p}=\Gamma_{p}^{\ast
}=C^{-1/3}\Gamma_{p}^{\prime}$ was optimized as in (\ref{gammasopt}). The
optimal squeezing is found to occur for $\Gamma_{p}^{\ast}=5.5\gamma_{0}$ and
$\Gamma_{p}^{^{\prime}\ast}=25\gamma_{0}$.}%
\label{Fig. 3}%
\end{center}
\end{figure}

The two pumping rates $\Gamma_{p}^{\prime}$ and $\Gamma_{p}$ can be optimized
in order to minimize $\Delta S_{\min}$. The optimal values of $\Gamma
_{p}^{\prime}$ and $\Gamma_{p}$ , denoted by $\Gamma_{p}^{\prime\ast}$ and
$\Gamma_{p}^{\ast}$, can be obtained easily from the minima in Fig.
\ref{Fig. 3},
which shows $\Delta S_{\min}$ versus $\Gamma_{p}^{\prime}$ for different
values of $\Gamma_{p}$. However, in order to better understand the behavior of
the variance with the cooperativity parameter, one can find approximate
expressions for $\Gamma_{p}^{\prime\ast}$ and $\Gamma_{p}^{\ast}$ in the
regime of interest $C\gg1$ and $\rho=\gamma_{0}/\kappa\ll1$ (typical
experimental values are $C\sim100-1000$, $\rho\sim1/2000$): explicitly, one obtains%

\begin{equation}
\Gamma_{p}^{\ast}\simeq\frac{\sqrt{3/2}\gamma_{0}}{\sqrt{\rho C}};\quad
\Gamma_{p}^{\prime\ast}\simeq\frac{\sqrt{3/2}\gamma_{0}}{\sqrt{\rho C^{1/3}}}.
\label{gammasopt}%
\end{equation}
Equations (\ref{gammasopt}) yield the following expression for the optimized
variance $\Delta S_{\min}^{\ast}$ in the regime considered,%

\begin{equation}
\Delta S_{\min}^{\ast}\simeq\frac{\lambda}{C^{1/3}}\quad\quad(C\gg
1,\ \rho=1/2000,\ \lambda\simeq1.74) \label{varopt}%
\end{equation}

\begin{figure}
[ptb]
\begin{center}
\includegraphics[
natheight=4.280000in,
natwidth=8.133600in,
height=3.5665in,
width=6.7568in
]%
{../../Paper 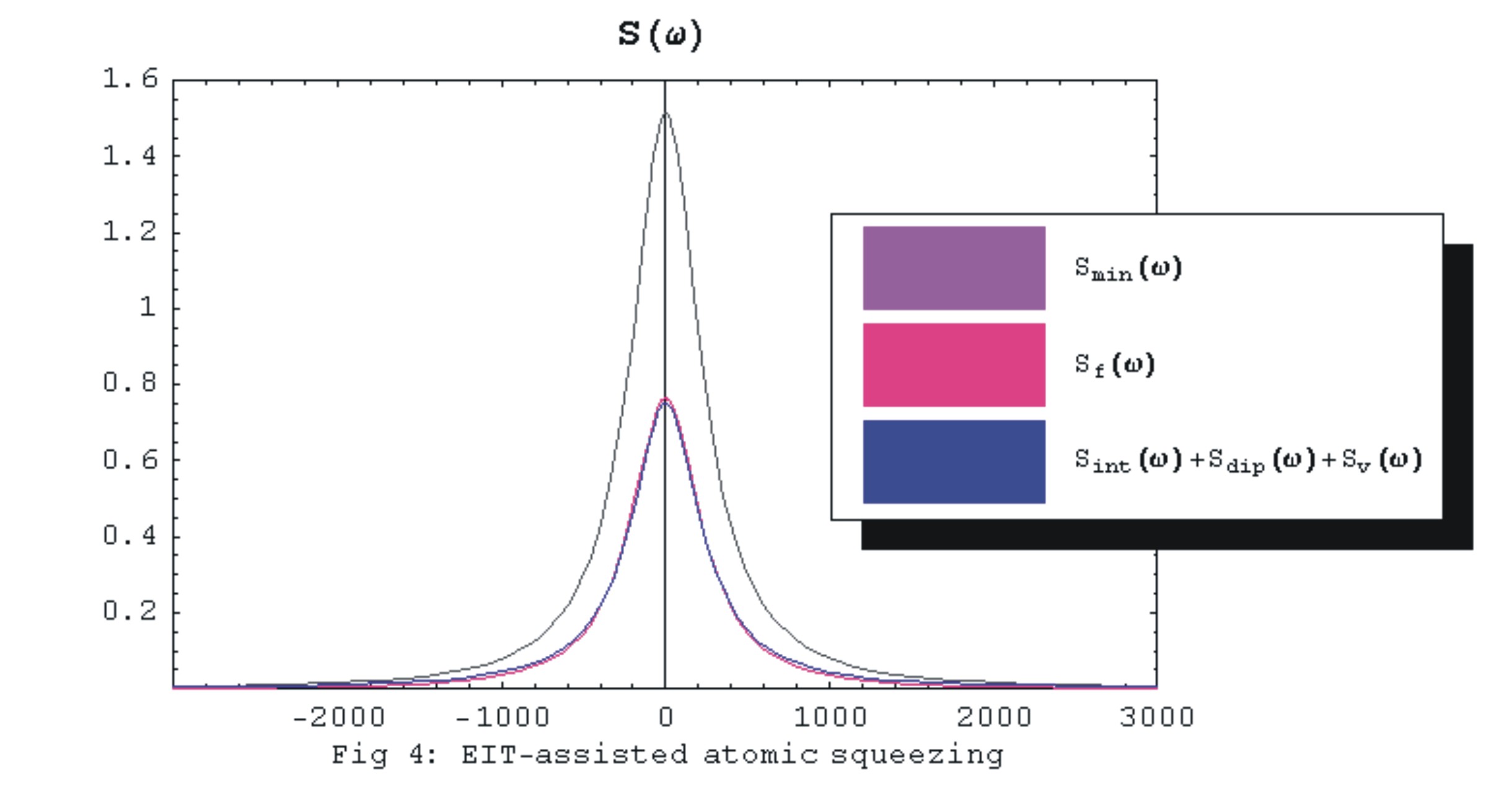}%
\caption{Minimal variance $\Delta S_{\min}$ versus $\Gamma_{p}^{\prime}$ (in
$\gamma_{0}$ units), for various values of the Raman pumping rate $\Gamma
_{p}=0.5,2,5.7,10,20$. $C$ and $\rho$ are equal to $100$ and $1/2000$. The
last curve shows the minimal variance in which $\Gamma_{p}=\Gamma_{p}^{\ast
}=C^{-1/3}\Gamma_{p}^{\prime}$ was optimized as in (\ref{gammasopt}). The
optimal squeezing is found to occur for $\Gamma_{p}^{\ast}=5.5\gamma_{0}$ and
$\Gamma_{p}^{^{\prime}\ast}=25\gamma_{0}$.}%
\label{Fig. 4}%
\end{center}
\end{figure}

The optimized atomic variance thus tends to $0$ with the cooperativity, and
could be made arbitrarily small with a large number of atoms. For Eq.
(\ref{varopt}) to be valid, $\Gamma_{p}^{\prime}$ and $\Gamma_{p}$ have to be
smaller than $\gamma^{\prime}$ and $\gamma$ for the adiabatic eliminations to
be justified. The optimized value of $\Gamma_{p}$ is typically of the order of
a few $\gamma_{0}$, so that the validity of our treatment is almost always
ensured under the optimized conditions: $\Gamma_{p}\sim\gamma_{0}$ and
$\gamma_{0}\ll\Gamma_{p}^{\prime}\ll\gamma^{\prime}$. The validity of the
approximations was checked with a full $4$-level calculation, the principle of
which has been described in Refs. \cite{Vernac,Vernac3}. We checked, in
particular, that the noise coming from the quantum fluctuations of the
EIT\ pumping fields is negligible in the regime of interest. It should be
noted that the convergence of $\Delta S_{\min}^{\ast}$ to $0$ with $C$ is
rather slow because of the exponent $1/3$; increasing the atoms by a factor
$10$ improves the squeezing by about $3$ dB. However, very good squeezing
values can be obtained for standard experimental conditions. In Fig.
\ref{Fig. 4}
the optimal squeezing in dB is plotted versus $C$. For $C=100$, we get $63\%$
($4.3$ dB) of squeezing. Increasing the cooperativity to $1000$
\cite{dalibard} would allow squeezing values of $83\%$ ($7.7$ dB).

\subsection{Contributions to the atomic noise spectrum}

\begin{figure}
[ptb]
\begin{center}
\includegraphics[
natheight=3.530200in,
natwidth=5.717300in,
height=3.5769in,
width=5.7752in
]%
{../../Paper 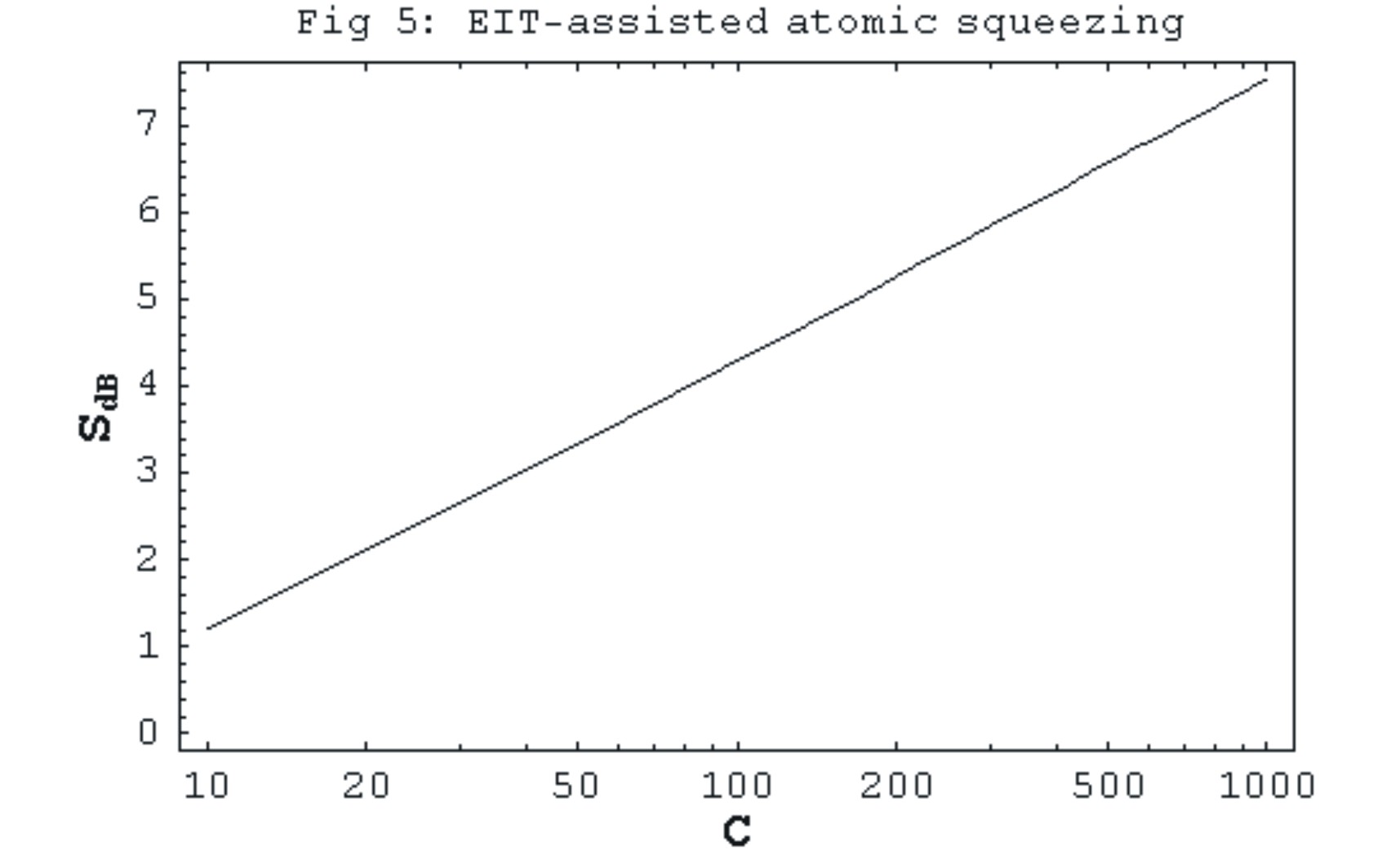}%
\caption{Contributions to the atomic noise spectrum: the contribution of the
fluctuations of the incident quantum field is of the same order of the three
other contributions. The spectrum width is $2\gamma_{+}\simeq455\gamma_{0}$ in
the case considered ($C=100$, $\rho=1/2000$, $\Gamma_{p}^{\ast}=5.5\gamma_{0}%
$, $\Gamma_{p}^{^{\prime}\ast}=25\gamma_{0}$).}%
\label{Fig. 5}%
\end{center}
\end{figure}

As in Ref. \cite{dantan}, it is interesting to go into the Fourier domain from
(\ref{deltaSx})-(\ref{deltaEq}) and plot the contributions to the atomic
spectrum. The atomic spectrum of the minimal component, $\left\langle \delta
S_{y}^{2}(\omega)\right\rangle $, is the sum of the incident field
fluctuations $S_{f}$ and the atomic noise $S_{at}$:%

\begin{equation}
S_{f}=\frac{N\rho C\Gamma_{p}^{3}}{\tilde{\gamma}_{0}^{3}}\frac{1}{\left\vert
D(\bar{\omega})\right\vert ^{2}};\quad S_{at}=\frac{N\rho\tilde{\gamma}_{0}%
}{2}\frac{1+\bar{\omega}^{2}}{\left\vert D(\bar{\omega})\right\vert ^{2}%
}\label{contributions}%
\end{equation}
where $\bar{\omega}=\omega/\kappa$ and $D(\bar{\omega})=(1-i\bar{\omega
})(\tilde{\rho}-i\bar{\omega})+2\rho C\Gamma_{p}^{2}/\gamma_{0}\tilde{\gamma
}_{0}$. These contributions are plotted versus frequency in Fig. \ref{Fig. 5}%
for optimized pumping values. Unlike the spectra derived in \cite{dantan}, we
see that in the present case both contributions are of the same order of
magnitude. The frequency width of the spectra can be found by noticing that,
if we are in the bad-cavity limit: $\kappa\gg\tilde{\gamma}_{0}$, and one
eliminates the field fluctuations in (\ref{deltaEp}) and rewrites
(\ref{deltaSy}), the effective time constant for $\delta S_{y}$ is $\gamma
_{+}=\tilde{\gamma}_{0}+2C\Gamma_{p}^{2}/\tilde{\gamma}_{0}$, which can be
made much greater than $\tilde{\gamma}_{0}$ (as in Fig. 5).

\subsection{Variation with the quantum cavity field intensity\label{field}}

If the cavity field mean value is non-zero, the fluctuations of $S_{y}$ are
coupled to both $E_{P}$ and $E_{Q}$ and Eq. (\ref{deltaSy}) has an additional
term proportional to $\left\langle A_{2}\right\rangle \delta S_{z}$. As a
consequence, $\delta S_{y}$ is coupled to the other spin components. The
minimal component in the plane orthogonal to the mean spin is then shifted,
and the shift increases when $\left\vert \left\langle A_{2}\right\rangle
\right\vert $ increases and its fluctuations are greater. Figure 6 shows the
minimal variance versus the cavity field mean value for a given set of
parameters. The squeezing is indeed destroyed when the field amplitude becomes
too large. Yet, there exists a substantial range of intracavity intensities
that do not destroy the squeezing too much. A lower limit for the field
amplitude is set by looking at the linearized equation for $\delta S_{y}$, in
which the field mean value is non zero and the field fluctuations have been
adiabatically eliminated%

\begin{equation}
\frac{d}{dt}\delta S_{y}=-\gamma_{+}\delta S_{y}+2\tilde{g}\left\langle
S_{z}\right\rangle \delta E_{P}+2\tilde{g}\left\langle A_{2}\right\rangle
\delta S_{z}+F_{y}^{\prime} \label{fluctuationsSy2}%
\end{equation}

\begin{figure}
[ptb]
\begin{center}
\includegraphics[
natheight=4.512600in,
natwidth=7.303300in,
height=3.7853in,
width=6.1099in
]%
{../../Paper 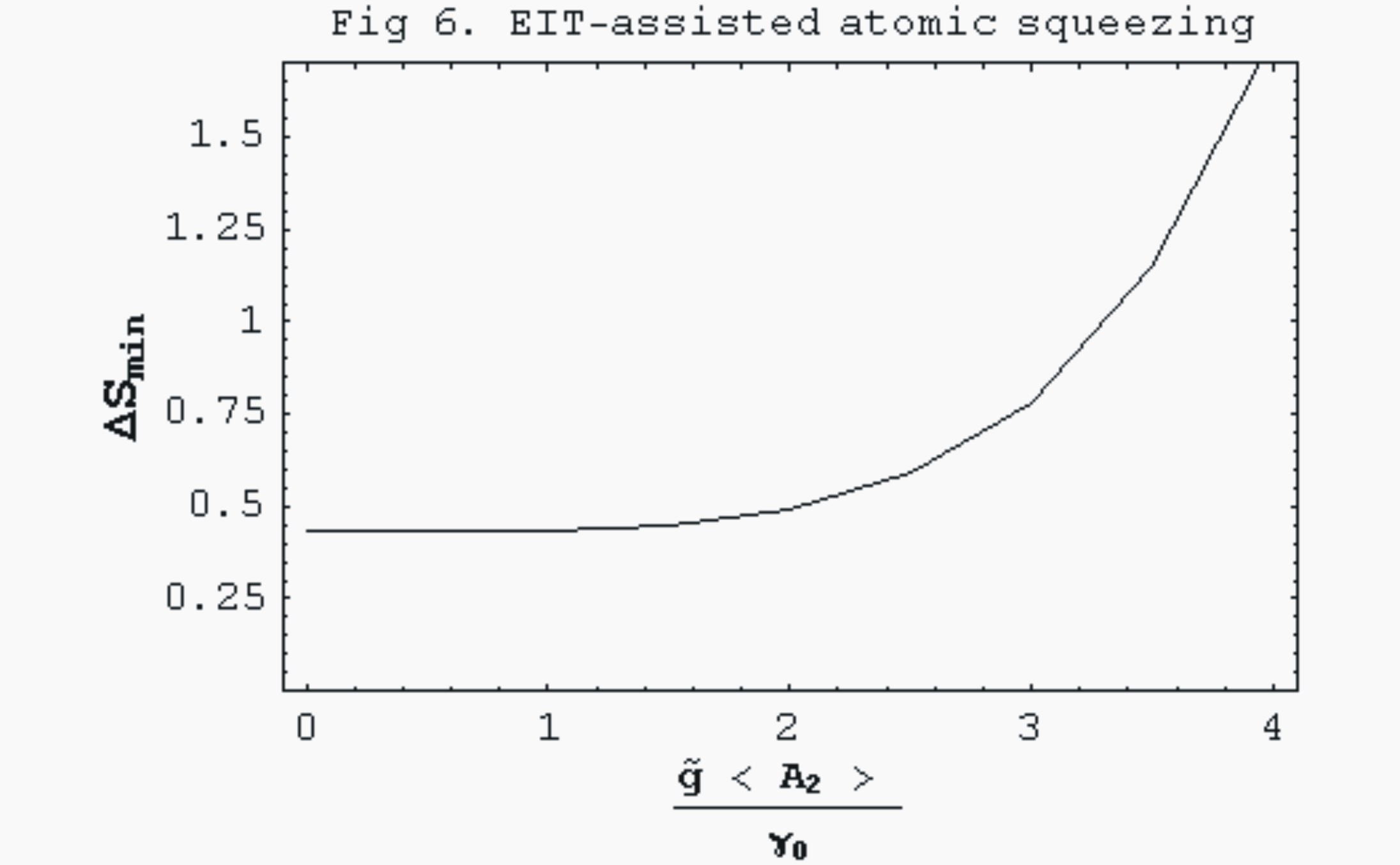}%
\caption{ Minimal variance $\Delta S_{\min}$ versus intracavity field
amplitude $\left\vert \tilde{g}\left\langle A_{2}\right\rangle /\gamma
_{0}\right\vert $. The parameters are $C=100$, $\rho=1/2000$, $\Gamma
_{p}=\gamma_{0}$, $\Gamma_{p}^{\prime}=4\gamma_{0}$. The field amplitude limit
as calculated from Sec. \ref{field} is $1.1\gamma_{0}$ in this case.}%
\label{Fig. 6}%
\end{center}
\end{figure}

Squeezing is degraded when $\gamma_{+}^{2}\left\langle \delta S_{y}%
^{2}\right\rangle \sim4\left(  \tilde{g}\left\langle A_{2}\right\rangle
\right)  ^{2}\left\langle \delta S_{z}^{2}\right\rangle $. Taking the
analytical expressions for the variances (not reproduced here for simplicity),
one thus gets a limiting value for the field amplitude $\left\vert \tilde
{g}\left\langle A_{2}\right\rangle \right\vert $. The calculations of Sec.
\ref{variance} for $\left\langle A_{2}\right\rangle =0$ are valid at least up
to this limit, as can be seen from Fig. \ref{Fig. 6}.

\section{Conclusion}

In \cite{Vernac,Vernac3,dantan}, the squeezing arose from the non-linear
interaction between the fields and atoms in the vicinity of the lower turning
point of the bistability curve. The atom-field system exhibits the analog of a
first-order phase transition and that quantum fluctuations are important near
the bistable point, allowing for either the field or the atoms to be squeezed.
The critical parameter governing the behavior is the cooperativity parameter
$C$. The underlying mechanism responsible for spin squeezing considered in
this paper is rather different from bistability squeezing. Rather than
originating in a regime where the amplitude of the fluctuations is either big
or small, the squeezing can be traced to a region of parameter space where the
fluctuations can be made small owing to the cooperative behavior of the atoms
due to the cavity coupling. At the same time, the spin mean value increases
faster than the fluctuations with increasing EIT pumping rate. We can then say
that the spin is pumped owing to the EIT\ interaction, while the fluctuations
are kept low by the cavity coupling. The consequence of this novel effect is
that the atomic squeezing no longer saturates at some constant value when one
increases the number of atoms. We would like to point out that the origin of
squeezing is rather complex, in the sense that both the cavity coupling and
the EIT pumping are necessary to produce squeezing, although each scheme,
taken alone, does not yield squeezing; hence our appellation of EIT-assisted
atomic squeezing. Note that the squeezing can be easily controlled via the EIT
intensity, and the optimization provided by the rather simple analytical
results of the effective system. In addition to the advantages of a cw
experiment, we would like to point out that the long life-time of the ground
state should render easier the squeezing detection and control.

\section{Acknowledgments}

The work of PRB was supported by the National Science Foundation under Grant
No. PHY-0098016 and the FOCUS Center grant, and by the U. S. Army Research
Office under Grant No. DAAD19-00-1-0412.

\section{Appendix}

We give the expression of the atomic diffusion coefficients in the case
discussed in Sec. \ref{diffusion}, when $\left\langle A_{2}\right\rangle =0$,
$\tilde{\delta}=0$, $\Delta_{c}=0$. They were evaluated with the Einstein
generalized relations \cite{Cohen} and grouped in the atomic diffusion matrix
$\left[  D_{at}\right]  $%

\begin{equation}
\left[  D_{at}\right]  =N\vspace{0.6cm}\left[
\begin{array}
[c]{ccc}%
\tilde{\gamma}_{0}+\Gamma_{p}-\frac{\Gamma_{p}^{\prime2}}{2\tilde{\gamma}_{0}}
& -\frac{\Gamma_{p}^{\prime2}}{2\tilde{\gamma}_{0}} & (\tilde{\gamma}%
_{0}+\Gamma_{p})\frac{\Gamma_{p}^{\prime}}{2\tilde{\gamma}_{0}}\\
-\frac{\Gamma_{p}^{\prime2}}{2\tilde{\gamma}_{0}} & \tilde{\gamma}_{0}%
-\Gamma_{p}-\frac{\Gamma_{p}^{\prime2}}{2\tilde{\gamma}_{0}} & (-\tilde
{\gamma}_{0}+\Gamma_{p})\frac{\Gamma_{p}^{\prime}}{2\tilde{\gamma}_{0}}\\
(\tilde{\gamma}_{0}+\Gamma_{p})\frac{\Gamma_{p}^{\prime}}{2\tilde{\gamma}_{0}}
& (-\tilde{\gamma}_{0}+\Gamma_{p})\frac{\Gamma_{p}^{\prime}}{2\tilde{\gamma
}_{0}} & \frac{\tilde{\gamma}_{0}}{2}-\frac{\Gamma_{p}^{\prime2}}%
{2\tilde{\gamma}_{0}}%
\end{array}
\right]  \label{Dat}%
\end{equation}

\end{document}